\documentclass[a4paper,11pt]{article}
\usepackage{jheppubnew} 
\usepackage{ifpdf}
\usepackage{tikz}
\usepackage{graphicx}
\usepackage{natbib}
\usepackage{amsmath,amssymb,amsfonts,bm,empheq,pgfplots}
\usepackage{hyperref}
\pgfplotsset{compat=1.18} 
\usepackage{subcaption}
\usepackage{nicefrac}
\usepackage[enableskew,vcentermath]{youngtab}
\usepackage{xcolor}
\usepackage{epstopdf}
\usepackage{url}
\usepackage{float}
\usepackage{slashed,framed,xcolor,tikz}
\usepackage{youngtab}

\newcommand{\be}{\begin{equation}}
\newcommand{\ee}{\end{equation}}

\usetikzlibrary{shapes.geometric, arrows}
\tikzstyle{startstop} = [rectangle, rounded corners,minimum width=2cm, minimum height=1cm,text centered, draw=black, fill=red!30, text width = 2cm]
\tikzstyle{eq} = [rectangle, minimum width=3cm, minimum height=1cm, text centered, draw=black, fill=orange!30]
\tikzstyle{arrow} = [thick,->,>=stealth]

\title{Spectral solutions for the Schr\"{o}dinger equation with a regular singularity}

\author[a]{Pushkar Mohile,}
 \author[b]{Ayaz Ahmed,}
\author[c]{T.R.Vishnu,}
 \author[b]{Pichai Ramadevi}
 \affiliation[a]{Department of Physics, Stonybrook, USA}
 \affiliation[b]{Department of Physics, Indian Institute of Technology Bombay, India}
 \affiliation[c]{Department of Physics, Raman Research Institute, 
Bangalore, India.}

\emailAdd{ramadevi@phy.iitb.ac.in}

\abstract{
We propose a modification in the Bethe-like ansatz  to reproduce  the hydrogen atom spectrum and the wave functions. Such a proposal provided a clue to attempt the exact quantization conditions (EQC) for the quantum periods associated with potentials $V(x)$ which are singular at the origin. In a suitable limit of the parameters, the potential can be mapped to $|x|$ potential. We validate our EQC proposition by numerically computing the Voros spectrum and
matching it with the true spectrum for $|x|$ potential. Thus we have given a route to obtain the spectral solution for the one dimensional Schr\"{o}dinger equation involving potentials with regular singularity at the origin.}

\begin{document}
\maketitle

\flushbottom

\section{Introduction}
The Bethe ansatz is one of the most powerful tools in the study of quantum integrable systems. It has profound applications in integrable spin chains. In fact, such an ansatz enables diagonalisation of Hamiltonian and obtains energy spectrum using simple algebraic arguments \cite{bethemetal,baxter82}. 

Even though the Bethe ansatz has been studied extensively in the context of integrable systems, there are interesting features that can be applied to other models.  One such striking feature is to compute the quantum spectrum from the classical limit of the integrable model \cite{pedrobethe}. Such calculations are based on the asymptotes of a set of polynomial equations, which we get from the  ansatz, called the Bethe equations. A nice review with applications to integrable quantum field theories and spin systems can be found in \cite{levkovich2016bethe}. In fact, this review article and \cite{pedrobethe} illustrate the exact energy spectrum of one-dimensional quantum harmonic oscillator (QHO) from a Bethe-like ansatz.  We believe such a neat concise Bethe-like ansatz approach must be generalisable for  other quantum mechanical systems. Hence, we investigated this approach for the hydrogen atom and proposed a modification in the Bethe-like ansatz. Interestingly, we succeeded in reproducing the energy spectrum and the corresponding wave functions.

Well behaved nature of wave function in quantum mechanics forces that the Bethe-like ansatz  for pseudo-momentum $p_0(x)$ (whose asymptotic behaviour in $\hbar \rightarrow 0$ matches classical momentum $p_{cl}(x)$) must have only simple poles. Unfortunately, we obtain higher order poles for any general polynomial potential of degree greater than 2. Hence, the Bethe-like approach fails for such potentials.  

 The natural extension of the Bethe-Like ansatz should be the exact WKB (Wentzel-Kramers-Brillouin) method \cite{voros1999airy,vorosreturn}  and the `thermodynamic Bethe Ansatz' (TBA) equations governing the quantum WKB periods \cite{DoreyTateo2007}. In fact, this route led to the energy spectrum for monic potentials \cite{DoreyTateo2007} and general polynomial potentials \cite{ito2019tba}. TBA involves  Borel transform and Borel re-summation techniques \cite{vorosreturn,ito2019tba} to handle diverging series as well as capture the singularities in the Borel plane. In fact, these discontinuities in the Borel transform encode information about other perturbative series  associated with different classical configurations \cite{delapham,iwaki2014exact}.
 
For QHO and hydrogen atom, no new information is obtained using TBA approach. However, for higher order polynomial potentials, we can capture the information about the other zeros of the potential from the singularity structure of the Borel re-summed function. TBA approach  is definitely powerful in computing the quantum periods for general polynomial potentials $V(x) = \sum_n a_n x^n$ \cite{ito2019tba}. Further, the exact WKB method advocated by Voros-Silverstone leads to an exact quantization condition (EQC) obeyed by the quantum periods \cite{ito2019tba,emery2021tba,marino2021advanced}. Thus the spectral solutions for any general polynomial potentials can be obtained. They have been numerically presented for the polynomial potential with a suitable choice of parameters $\{a_n\}$ \cite{ito2019tba,emery2021tba}.

For other potentials with a simple pole and a double pole at the origin, the modification in the TBA analysis has been systematically elaborated in \cite{ito2020tba}. However, due to the singularity at the origin, the EQC is still an open problem. The Bohr-Sommerfeld quantization to obtain the spectral solution is not correct. 

The main theme of this paper is to propose a correction to the quantum period near the singular origin to modify the existing polynomial potential EQC. Such a proposal is motivated from our Bethe-like ansatz for the hydrogen atom. We validate our EQC proposition through an example whose energy spectrum is known.

We know that the wave functions for the $|x|$ potential
are the Airy functions. In fact, the zeros of the Airy function and its derivatives give the true spectrum \cite{airyschwinger,englert2020particle}. We performed a naive TBA approach for the $|x|$ potential with two turning points and obtained the spectrum using the EQC of the QHO.  Our calculated spectrum did not match the true spectrum for the low lying energy states. This exercise indicates that the conventional (TBA \& EQC) approach, of finding spectral solutions for polynomial potentials, cannot be applied to the potentials with a derivative singularity at the origin. 

Incidentally, the solutions  for  $|x|^{2n+1}$ potentials using spectral determinant approach are discussed in \cite{voros1999airy,suzuki2015elementary}. However, our aim was to modify the polynomial potential TBA as well as the EQC to reproduce the true spectrum for these potentials.

As a first step, we showed that the $|x|$ potential can be viewed as the potential with a simple and a double pole \cite{ito2020tba} for a suitable limit of parameters. With this choice of parameters, we numerically computed the quantum periods using TBA equation. Then, using our EQC proposition we obtained the Voros spectrum. In fact, our numerical results for the Voros spectrum match well with the true spectrum. Our validation for $|x|$ spectrum reinforces that the proposed EQC is applicable for the potentials with a regular singularity \cite{ito2020tba}.

  The plan of the paper is as follows: In section \ref{sec:BetheLike}, we briefly review Bethe-like ansatz and present the  spectrum of QHO. Then we propose a modification in the Bethe-like ansatz necessary to reproduce the hydrogen atom spectrum. In section \ref{sec:wkbmethod}, we have discussed exact WKB method and Borel resummation technique to deal with divergent perturbative series. We summarise the salient details of TBA in section \ref{gensec:TBA} with some simple potentials as illustrative examples. We discuss the  \(|x|\) potential and its relation to potentials with regular singularity in \ref{sec:TBAmodx}. In section \ref{sec:EQC}, we focus on our proposition of EQC for potentials, singular at the origin. We summarise and present some of the open problems in the concluding section \ref{sec:discussion}.

\section{Bethe-Like Ansatz \label{sec:BetheLike}}
For completeness and clarity, we will first review the salient features of Bethe-like ansatz approach for the quantum harmonic oscillator (QHO) spectrum \cite{pedrobethe,levkovich2016bethe}. Then, we present our proposal of modified Bethe-ansatz giving hydrogen atom spectrum.

 For QHO, a set of Bethe-like equations can be written for the roots of the wave function. This relies on the nonlinear transformation of time independent Schr\"{o}dinger equation (TISE) 
 \begin{equation}
	-\frac{\hbar^2}{2m}\frac{d^2}{dx^2} \psi(x) + V(x)\psi(x) = E\psi(x),
	\label{e:TISE}
	\end{equation}
  into Riccati equation:
  \begin{equation}
	p^2 - i\hbar p' = 2m(E-V),
	\label{eq:Ricatti}
	\end{equation}
where
\begin{equation}
p(x) = \frac{\hbar}{i} \frac{\psi'(x)}{\psi(x)}.
	\label{eq:ColeHopf}
\end{equation} 
Note that  \(p(x)\)(\ref{eq:ColeHopf}) has singularities at the zeros of the wave function \(\psi(x) \). Such singularities are handled by doing an analytic continuation of \(p(x) \rightarrow p(z)\) in the complex plane. The nature of {\it complex function} \(p(z)\) can be fixed from the generic behaviour of the wave function \(\psi(x) \),  i.e., \(\psi(x)\) must be normalisable.  Suppose
we  allow  second or higher order poles  for  $p(z)$, 
\begin{equation}
p(z) \propto (z-a)^{-n} ~{\rm for} ~n \geq 2.
\end{equation} 
Then the wave function 
$$ \psi(z) \propto \exp\left(\int p(z) dz\right) = \exp[-n(z-a)^{-n+1}],$$ 
 has essential singularities. This implies that \(p(z)\) can have at most simple poles. Note that the roots of the bound state wave functions \(\psi(x) \)  are discrete and isolated \cite{berezin2012schrodinger}.  
 
 For highly excited states, we can take  the classical limit $\hbar \rightarrow 0$.  Clearly, 
\(p(z)\) in the classical limit 
 \begin{equation}
\lim_{\hbar \rightarrow 0}~ p(z) \equiv p_{\rm cl}(z) = \pm \sqrt{2m(E-V(z))},
\label{classicalp}
\end{equation}
denotes the familiar classical momentum of the particle which has  branch cut singularity. It must be puzzling as to where from this branch cut emerges in the classical limit. It can only be formed when the discrete poles present in \(p(z)\) `condense' to a continuum as we approach the classical limit. Hence, we can conclude that the  poles condense into the branch cut in the classical limit. Note that \(p(z)\) is applicable for classically allowed region ($E \geq V(x)$) as well as classically forbidden region ($E < V(x)$). Hence \(p(z)\) is referred to as pseudo-momentum.

 We will now review QHO spectrum from  the Riccati equation to see the resemblance with Bethe ansatz equations.

 Let us examine the classical limit \(p_{\rm cl}(z)\)(\ref{classicalp}) for QHO, of mass $m$ and angular frequency $\omega$ ,  whose $V(x) = V_{\rm QHO} = m\omega^2 x^2/2.$  The function (\ref{classicalp}) has a square root type branch cut, with branch points at the two turning points 
$$ z = \pm \sqrt{2E/m \omega^2}.$$
Our aim is to determine the allowed energy eigenvalues $E$ for QHO. In order to achieve this,  we probe the asymptotic behaviour of \(p(z)\) as \(z \to +\infty\) on the real axis :
$$p \sim im \omega z + \mathcal O({\frac{1}{z}})~~{\rm and }~ ~ p' \sim  im \omega+ \mathcal O({\frac{1}{z^2}}).$$
Notice that the leading term in asymptotic $p^{\prime}(z)$ is a constant and must be included so that Riccati equation gives
\begin{eqnarray} 
\lim_{z \rightarrow \infty} p(z) \equiv  p_o(z) &=&\lim_{z \rightarrow \infty} \sqrt{2m\left[ (E - \frac{\hbar \omega}{2}) - \frac{m \omega^2 z^2}{2} \right]} \nonumber\\
~& \sim& im\omega z { - i\frac{(E- \hbar  \omega/2)}{\omega z} + \mathcal{O}(1/z^3), } 
\end{eqnarray} 
where $\lim_{\hbar \rightarrow 0} p_o(z) = p_{\rm cl} (z)$(\ref{classicalp}). Notice that the asymptotic behaviour of $p_o(z)$ is almost like the classical momentum(\ref{classicalp}),  if we shift 
\begin{equation}
E \rightarrow E-{\frac{1}{2} } \hbar \omega.
\end{equation}
In fact, the branch cut of $p_o(z)$  includes all the poles of $p(z)$.  It is exciting to see the natural emergence of  quantum shift in the energy by  $\hbar \omega/2$ from the Riccati equation for QHO. The asymptotic behaviour of \(p(z)\equiv p_o(z)\), which has a branch cut, is due to the condensation of simple poles of $p(z)$. This leads to the following Bethe-like ansatz for \(p(z)\) having $N$-simple poles:
	\begin{equation}
	p(z) = im\omega z + \frac{\hbar}{i} \sum_j^N \frac{1}{z - z_j}. 
	\label{eq:bethelikeqho}
	\end{equation}
Here, we make the choice of sign in the leading term (\(im\omega z\))  so that the wave function remains normalisable. The set \(\{z_j\}\) corresponds to the \(N\) roots arising from the nodes of the \(N^{th}\) excited eigenfunction. Incorporating the  key observation of the Bethe-like approach,  the contour integration around the branch cut in \(p_o(z) \) must give the residues due to the simple poles of $p(z)$(\ref{eq:bethelikeqho}):
\begin{equation}
	\oint_{\gamma} \sqrt{2m\left[ \left(E- \frac{\hbar\omega}{2} \right) -V(z) \right]} dz = \sum _j ^N \text{Res}_{x_j} = 2\pi \hbar N,  \label{contour}
	\end{equation}
where \(N\) is the number of roots for the \(N^{th}\) excited state  and \(\gamma\) is a contour around the branch cut.  By doing this contour integral, we get 	
	\begin{equation}
	2\pi \left(E- \frac{\hbar \omega}{2}\right) = 2 \pi (N) \hbar \omega, 
	\end{equation}
which gives us the energy spectrum of the QHO:
	\begin{equation}
	E = \left( N+ \frac{1}{2} \right) \hbar \omega.
	\end{equation}
We have to deduce the  wave functions $\psi_N(x)$ corresponding to the $N$-th excited energy level from the Riccati equation.
When we substitute the ansatz(\ref{eq:bethelikeqho}) into the Riccati equation(\ref{eq:Ricatti}) 
and equate the coefficients of each of the terms \(1/(z-z_j)\)  to zero,
we obtain Bethe-like equations for the the roots \(\{z_j, j=0,1,\ldots N\}\) of the $N$-th excited state wavefunction $\psi_N(x)$ :
\begin{equation}
	z_j = \frac{\hbar}{m\omega} \sum_{i\neq j} \frac{1}{z_j - z_i}, \quad \forall j = 1,2,3,..N. 
	\label{eq:bethelikeeqsQHO}
	\end{equation}
This system of polynomial equations is solvable, with solutions to \(z_j\) being the roots of Hermite polynomials when the factor \(\hbar/m\omega \) is scaled to \(1\). We have solved  this system of polynomial equations for $N \leq  3$ using Mathematica and tabulated (see Table \ref{fig:polytableQHO}). These roots computed are matching with the roots of Hermite polynomials.
	\begin{table}
    	\centering    
	\begin{tabular} {|c|c|c|}
        \hline 
        \(N\) & Equations & Solutions \\ 
        \hline
        0 & No equations  & No roots  \\
        \hline   
        1 & \(x_1 = 0\) & \(x_1 = 0\) \\
        \hline
        2 & \(x_1(x_1 - x_2) =1\) & \(x_1 = -1/\sqrt{2}\)\\ 
        \phantom{1} & \(x_2(x_2 - x_1) = 1\)  & \(x_2 = 1/\sqrt{2} \) \\ 
       	\hline
        3& \(x_1(x_1-x_2)(x_1 - x_3) = 2x_1 - x_2 - x_3  \) & \(x_1 = -\sqrt{3/2}\)\\ 
        \phantom{1}& \(x_2(x_2-x_1)(x_2-x_3) = 2x_2 - x_1 - x_3\) & \(x_2 = 0\)\\
        \phantom{1} & \(x_3(x_3 - x_1)(x_3-x_2) = 2x_3 - x_1 - x_2 \) & \(x_3 = \sqrt{3/2}\) \\ 
        \hline
       	\end{tabular}
        \caption{Bethe equations for QHO roots for \(\frac{\hbar}{2m\omega}=1\).}
        \label{fig:polytableQHO}
	\end{table}
	Once we know the roots $\{x_j\}$'s  for any $N$,  the corresponding energy eigenfunction is constructed as 
	\begin{equation}
	\psi_N(x) = \exp [\int p(x)dx ] = A \exp\left(\frac{-x^2}{2}\right)  \prod_{j =1}^N(x-x_j),
	\end{equation}
where \(A\) is determined by normalisation. 
The flowchart (Table \ref{flow: QHO}) gives a concise summary of the Bethe-like methodology for QHO.
Thus we have elaborated the powerfulness of this Bethe-like approach to obtain the complete QHO energy spectrum and the corresponding wave functions. Particularly, the branch cut in asymptotic behaviour of $p_o(z)$ is accountable by condensation of simple poles. Further, the well known zero point  energy  ($\hbar \omega/2$) appeared naturally.
It is not clear whether this methodology works for arbitrary potential  $V$. As a  first step in this direction, we have attempted hydrogen atom in the following subsection.
	
	 \begin{table}
	    \usetikzlibrary{shapes.geometric, arrows}
            \tikzstyle{startstop} = [rectangle, rounded corners, minimum width=3cm, minimum height=1cm,text centered, draw=black, fill=blue!20]
            \tikzstyle{io} = [trapezium, trapezium left angle=70, trapezium right angle=110, minimum width=3cm, minimum height=1cm, text centered, draw=black, fill=blue!30]
            \tikzstyle{process} = [rectangle, minimum width=3cm, minimum height=1cm, text centered, draw=black, fill=green!20]
            \tikzstyle{decision} = [diamond, minimum width=3cm, minimum height=1cm, text centered, draw=black, fill=green!30]
            \tikzstyle{arrow} = [thick,->,>=stealth]       
                \begin{tikzpicture}[node distance=2cm]
                \small {        
             \node (start) [startstop] [xshift=0.5cm, yshift=-0.4cm,align=center]
            {\(p^2 - i\hbar p' = 2m(E- \frac{1}{2} m \omega^2 x^2) \) };   
                \node (dec1) [startstop, right of=start, xshift=4.5cm, align=center] {\(p = im\omega x + \hbar/i \sum_j 1/(x-x_j)\)};
                \draw [arrow] node[anchor=south] {Riccati Equation} (start) --node[xshift=3.5cm,yshift=0.5cm, anchor=south] {Bethe-like ansatz}(dec1);                          
 \node (dec2) [process, below of=dec1, xshift=-4.5cm, yshift=-0.25cm,align=center] {\(p_{cl} = \sqrt{2m[E- \frac{1}{2} (m\omega^2 x^2 )]} ]\)};
                \draw [arrow] (dec1) -- node[xshift=0.2cm,anchor=west] {\(\hbar \rightarrow 0\)} 
                node[anchor=east] {classical limit}(dec2);               
\node (AsympMatching) [startstop, right of = dec2,xshift=-1.cm,yshift=-2.5cm]  {\(im\omega x - i{[E-\frac{1}{2}( \hbar \omega)] /\omega x}+ \mathcal{O}(1/x^3) = im\omega x + \hbar/i \sum 1 / (x - x_j)  \) };
\node (AsympMatch)[startstop, xshift=-2.5cm,below  of =AsympMatching] {\(z_j = \frac{\hbar}{m\omega} \sum_{i\neq j} \frac{1}{z_j - z_i}, \quad \forall j = 1,2,3,..N.  \)};
 \draw [arrow] (dec2) -- node[xshift=-0.1cm,yshift=-0.1cm, anchor=east] {(Poles to)}node[xshift=0.2cm,yshift=-0.1cm, anchor=west]{ (Branch Cut)} node[anchor=south]{matching asymptotes }(AsympMatching);
 \draw [arrow](AsympMatching)--node[anchor=west]{Bethe equations}(AsympMatch);
 \node(solutions)[process, xshift=6.1cm, right of=AsympMatch]{\(E_N= (N+{\frac{1}{2}}) \hbar \omega~,\psi_N(x)\propto H_N(x)\)};
 \draw[arrow](AsympMatch)--node[anchor=south]{Spectrum}(solutions);}
 \end{tikzpicture} 
 \caption{Flowchart for QHO spectrum from  Bethe-Like approach}
\label{flow: QHO}   
\end{table}                     
\subsection{Bethe-Like Ansatz for the Hydrogen Atom \label{sec:HABethe}}
For the hydrogen atom, the  potential energy is $V(r) \propto {\frac{1}{r}}$. Clearly, rotations in the three-dimensional space leaves the Hamiltonian of the hydrogen atom invariant. Even though it is a three-dimensional system, we can view the hydrogen atom as an effective one dimensional system in radial coordinate $r$. By rewriting the radial part $R_n^{l}(r)$ of the wavefunction 
$\psi_{n,l,m}(r,\theta,\phi)= R_n^{l} (r) Y_{l m} (\theta, \phi)$  as 
$u_n^{l}(r) = r R_n^{l} (r)$, it is easy to check that the radial part of the equation resembles the one-dimensional Schr\"{o}dinger equation for $u_n^{l} (r)$  with effective potential energy
\begin{equation} 
V_{eff} (r) = -\frac{e^2r} {4\pi \epsilon_o r} + \frac{ \hbar^2 l(l+1)} {2m r^2 },
\end{equation}
where the quantum number $l$ refers to the orbital angular momentum. Following the arguments in the previous section,  the pseudo-momentum $p(r)$ is a rational function with simple poles at \(\{r_j\}\) for regular functions $u_n^{l}(r)$ having zeros at the points \(\{r_j\}\) where \( j=1,2, \ldots N.\)  The residues of \(p(r) \) at these poles must be $2\pi i \times {\hbar}/{i}.$ We should keep in mind the following  key differences between the  harmonic oscillator and the hydrogen atom:\\ 
(i)  The passage from the three-dimensional problem to the effective one-dimensional system should introduce an additional pole at  \(r=0\). \\
(ii)  the domain of definition is \(r\geq 0\).  \\
(iii)  The asymptotic form of pseudo-momentum $p(r)$ matches exactly the asymptotic values of the classical momentum $p_{cl}(r)$.\\
Hence, for highly excited states, $p(r)$ is the classical momentum (there is no zero point energy shift). In the classical limit, we observe the poles at \(\{r_j\}\) condense to form a square root branch cut of $p(r)$.  At large values of \(r\), \(p \to -\sqrt{2mE}\) as \(V \to 0\), where the negative branch of the square root is chosen to prevent the wave function $u(r)$  from blowing up at \(\infty\). We will now  focus on the energy and the wave function for the $s$-orbitals of hydrogen atom ($l=0$) which will give us the insight to generalise the Bethe-like ansatz for $l \neq 0$.

\subsubsection{Spectrum for $l=0$}
Let us propose an ansatz for $p(r)$ for the $s$-orbitals whose orbital angular momentum $l=0$ to obtain the energy spectrum and the corresponding wave function $u(r)$. Incorporating the  asymptotic form  of $p(r)$ and its  poles at \(\{r_j\}\), we propose the following Bethe-like ansatz.\\
{\bf Proposition 1:}
	\begin{equation}
	p = -\sqrt{2mE} + \frac{\hbar}{i} \frac{1}{r} + \frac{\hbar}{i}\sum_{j=1}^N\frac{1}{r-r_j}.
	\label{eq:bethelike}
	\end{equation}
with $N+1$ poles including the pole at $r=0$. Recall this additional pole was not there in the harmonic oscillator. The large \(r\) limit can be expressed as  
	\begin{equation}
	\lim_{r \rightarrow \infty} p(r)  = \lim_{r \rightarrow \infty} -\sqrt{2mE\left(1 -\frac{b}{rE} \right)} \sim
	-\sqrt{2mE} + \frac{b\sqrt{m}}{r\sqrt{2 E}} + \mathcal{O}\left(\frac{1}{r^2} \right),
	\end{equation}
	where \(b = e^2/4\pi\epsilon_0\).
Since the poles must condense to this branch cut, on doing a contour integration around the set of zeroes \(\{r_j\}\), the residues must equate on both sides giving us 
 	\begin{equation}
	\frac{b\sqrt{m}}{\sqrt{2 E}} = \frac{\hbar}{i} (N+1). \label{eq:hydenergy}
	\end{equation}
Here, $N$ is the number of zeros of the $s$ orbital wavefunction $u_n^{l=0} (r)$, and one more pole from \(r=0\) for $R_n^0(r)$. On rearranging and substituting \(b = e^2/4\pi\epsilon_0\) we get 
	\begin{equation}
	E_n = -\frac{e^4m}{32\hbar^2\pi^2\epsilon_0^2 n^2}= - \frac{e^2}{2 a_0} \frac{1}{n^2} ~{\rm where} ~n=N+1,
	\end{equation}
where $a_0= \hbar^2/ (m e^2)$ is the Bohr radius.
This matches exactly with the hydrogen atom energy spectrum. Further, (\ref{eq:bethelike}) allows us to fix the roots of the wave function $u(r)$  by requiring that the coefficients of each of the \({1}/{(r-r_i)}\) terms add up to zero  in the Riccati equation: 
	\begin{equation}
	\sqrt{2mE_n} = \frac{\hbar}{i} \frac{1}{r_i}+ \frac{\hbar}{i}\sum^N_{ \{k\neq i\} =1} \frac{1}{r_i-r_k} \phantom{1} \forall i \in 1,2,..N. 
	\label{eq:betheroots}
	\end{equation}
 This gives us a set of  Bethe-like equations  to solve and determine the roots $\{r_i\}$. 
 For \(N=1\),  we get \(r_1 = 2a_0\). We have tried to work out the roots for $N=n-1\leq 3$ using Mathematica and presented the results in Table \ref{fig:polytable} for $a_0=1$.  Once we have explicitly found the roots,  we can then integrate the ansatz for \(p\) to obtain the wave function $R_n^{l=0}(r)$. We see explicitly that for \(N\) zeros of the wave function $u(r)$ we get
	\begin{equation}
	R_n^0(r)  = A\exp({-\sqrt{2m|E_n|}~ r})L_n^0(r).
	\end{equation}
where $L_n^0(r)$ are  the Laguerre polynomials.
Here the negative branch of the square root is chosen to ensure \(R_n(r) \to 0\) as \(r\to \infty\) and \(A\) is  the  normalisation constant. \(L_n^0(r) = \prod_{i=1}^N(r-r_i)\) is a polynomial with roots at 
\(r_i\) found from Bethe-like equations.
 For \(N = 0\) and \(N = 1\), we get the wave functions 
	\begin{equation}
      	R_1^0(r)  = A_1\exp({-\frac{r}{a_0}}), ~~
	R_2^0(r) = A_2 \exp({-\frac{r}{2a_0}})(r-2a_0).
	\end{equation} 
Using the mathematica program, we can deduce $N=2$ for $a_0=1$ as
\begin{equation}
      	R_3^0(r) = A_3\exp (-\frac{r}{3})(r - 3/2(3-\sqrt{3}) ) (r - 3/2(3+\sqrt{3})). 
	  	\end{equation}
	\begin{table}
    	\centering    
      	\begin{tabular} {|c|c|c|}
        \hline 
        \(n = N+1\) & Equations & Solutions \\ 
        \hline
        1 & No equations  & No roots  \\
        \hline   
        2 & \(r_1 = 2\) & \(r_1 = 2\) \\
        \hline
        3 & \(r_1^2 -r_1r_2 -6r_1  + 3x_2 =0\) & \\ 
        \phantom{1} & \(r_2^2 - r_1r_2 - 6r_2 + 3r_1 =0\) & \(r_1 = {3/2}(3-\sqrt{3}) \) \\ 
        \phantom{1} & \(1/r_1 + 1/r_2 = 2/3\) &\( r_2 = {3/2}(3+\sqrt{3})\) \\
        \hline
        4& \(r_1(r_1-r_2)(r_1 - r_3) = 4 (3r_1^2 -2r_1r_2 - 2r_1r_3 + r_2r_3)  \) & \(r_1 = 1.871\)\\ 
        \phantom{1}& \(r_2(r_2-r_1)(r_2-r_3) = 4(3r_2^2 -2r_2r_1 - 2r_2r_3 + r_1r_3 )\) & \(r_2 = 6.618\)\\
        \phantom{1} & \(r_3(r_3 - r_1)(r_3-r_2) = 4( 3r_3^2 -2r_3r_1 - 2r_3r_2 + r_1r_2 ) \) & \(r_3 = 15.517\) \\ 
        \phantom{1} & \(1/r_1 + 1/ r_2 + 1/r_3 = 3/4\) &  \\
        \hline
       	\end{tabular}
        	\caption{Bethe equations for Hydrogen atom roots for \(l=0\) and \(a_0=1\)}
        	\label{fig:polytable}
    	\end{table}
	We  will  generalise the ansatz for $p(r)$ for arbitrary $l$ in the following subsection.
\subsubsection{Spectrum  for  \(l \neq 0\)}
{\bf Proposition 2:} Generalising \textit{proposition 1} in (\ref{eq:bethelike}), for any $l$, Bethe-like ansatz for $p(r)$   is
	\begin{equation}
      	p = -\sqrt{2mE} + \frac{\hbar}{i}\left[ \frac{l+1}{r} +\sum_j \frac{1}{r-r_j} \right].\label{eq:bethelike-withl}
  	\end{equation} 
Such an ansatz  will take care of the additional multiplicity of the root at $r=0$ for wave function \(R_n^{l} (r)\). Further, the  term \(l(l+1)/r^2\) term in the effective potential will be accounted for by the modified ansatz. 

The calculation of the energy spectrum $E_n$ for the modified ansatz is almost the same :
\begin{equation}
	E_n = -\frac{e^4m}{32\hbar^2\pi^2\epsilon_0^2 n^2}= - \frac{e^2}{2 a_0} \frac{1}{n^2} ~{\rm where} ~n=N+l +1.
 \label{betheansatzhydro}
	\end{equation}
Note that $n$ counts the total number of roots of the wave function $R_n^{l}(r)$ and 
$l+1$ counts the degeneracy of the root at the origin $r=0$. Interestingly, we observe the  bound on $l$ to be:
\begin{equation}
l \leq n.
\end{equation}
Substituting the modified ansatz in the Riccati equation and equating the coefficients of 
$1/(r_i-r_j)$ to zero we get the following set of Bethe equations for the roots:
	\begin{eqnarray}
    	\sum_j \frac{1}{r_j} &= &\frac{1}{a_0}\left(\frac{1}{l+1} - \frac{1}{N+l+1}\right),  \\
 	\frac{\hbar}{i} \sqrt{2mE_n} &=& \hbar^2 \sum _{i\neq j}\frac{1}{r_j-r_i} + \hbar^2 \frac{(l+1)}{r_j} ,  ~j = 1,2,..N. 
	\end{eqnarray}
Solving these equations for every $N$ will give the solutions for the roots leading us to write the  associated Laguerre polynomials:
\begin{equation}
R_n^{l} (r) \propto L_n^{l} (r).
\end{equation}
From  our   {\bf proposal} of reproducing  hydrogen atom spectrum, it is tempting to speculate whether  the spectrum for arbitrary  potential  $V(x) =\sum_k a_k x^{\pm k}$  can be elegantly obtained. Unfortunately, well behaved nature of wave function requiring $p(z)$ to have only simple poles is inconsistent with the asymptotic expansion of classical momentum $p_{cl}(x)$:
\begin{eqnarray}
  \lim_{ x\rightarrow \infty} \sqrt{2m (E- x^n)} &\sim& i\sqrt{2mx^n}  \left( 1 - \frac{E}{x^n}   + \mathcal{O}\left(\frac{1}{x^{2n}} \right) \right), \quad \text{for} \quad  n > 0,  \\
\lim_{x \rightarrow \infty} 	\sqrt{2m(E - x^{n})} &\sim& \sqrt (2mE) (1 - x^n/E + \mathcal{O}(x^2n)), \quad \text{for} \quad n < 0.
\end{eqnarray}
It appears that the Bethe-like ansatz requires the wave function to factorise into two parts:
	\begin{equation}
      	\psi(x) = f(x)g(x) ~;~p(x) = \frac{\hbar}{i} (f'/f + g'/g).
  	\end{equation}
Here \(f(x)\) governs the asymptotic behaviour of \(\psi(x)\) in the limit \(|x|\to \infty \) and \(g(x)\) is the polynomial that encodes the roots of \(\psi(x)\). In the Bethe-like ansatz, we assumed that the asymptotic behaviour of the wave function is governed only by the leading order asymptotic behavior, which looks like \(\exp[-x^2]\) for the QHO and \(\exp[-r/na_0]\) for the hydrogen atom.  This is the most trivial possible choice of the asymptotic behaviour of the function. Such a choice of asymptote does not appear for other potentials. We  may have more contributions to the asymptote due to non-perturbative corrections. Hence, we will have to go beyond Bethe-like ansatz to tackle spectral solution for higher degree polynomial potentials.

 \section{WKB  Method}
 \label{sec:wkbmethod}
\begin{figure}[ht]
 	\centering
 \begin{tikzpicture}
  \begin{axis}[
 		xmin = -2, xmax = 2,
 		ymin = -5, ymax = 5,
 	axis line style={draw=none},
tick style={draw=none},
 	yticklabels={,,},
 	xticklabels={,,}
 	]
 	\addplot[
 			domain = -2:2,
 			samples = 50,
 			smooth,
 			thick,
 			red,
 	] {0};
 	\addplot[
 			domain = -2:2,
 			samples = 50,
 			smooth,
 			thick,
 			black,
 		] {3*x^2-3};
 	\node at (axis cs:-1.2,0) {\(E\)};
 	\node at (axis cs:-1.8,1) {\(I\)};
 	\node at (axis cs:0,1) {\(II\)};
 	\node at (axis cs:1.8,1) {\(III\)};
 	\node at (axis cs:-1,-2.5) {\(q_1\)};
 	\node at (axis cs:1,-2.5) {\(q_2\)};
 	\node at (axis cs:0.2,-3.2) {\(\gamma\)};
 	\draw[gray] (axis cs:-1, 0) -- (axis cs:-1, -2.1) [dashed];
 	\draw[gray] (axis cs:1, 0) -- (axis cs:1, -2.1) [dashed];
 	\draw[rounded corners=0.2cm, thick] (axis cs:-1.1, -2.9) rectangle (axis cs:1.1, -2.1) {};
 \end{axis}
 \end{tikzpicture}
 \caption{WKB loops for potential \(V = 3x^2 \)}
 \label{fig: WKBperiodsQHO}
 \end{figure}
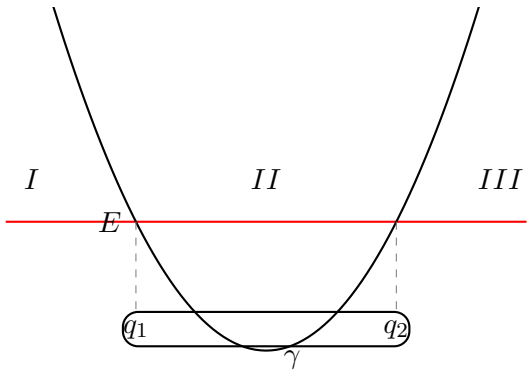

In the conventional  WKB (Wentzel-Kramers-Brillouin) approximation, we are familiar with the Bohr-quantization condition
\begin{equation}
\Pi_{\gamma,0} =\oint_{\gamma} p_{cl}(x)  dx =  2\pi N,
\end{equation}
which matches well for large $N$ excited states for any potential $V(x)$. $\Pi_{\gamma,0}$ is sometimes referred to as classical WKB period evaluated for a contour $\gamma$ around two  turning points $x_{\pm}$ where $V(x_{\pm})=E$. For QHO,
$x_{\pm} = \pm \sqrt{2mE}$ and curve $\gamma$ is indicated in Figure \ref{fig: WKBperiodsQHO}.
In order to find perturbative corrections to the spectrum,  we expand the pseudo-momentum $p(x)$ as power series in $\hbar$
\begin{equation}
p(x) =\sum_{n=0}^{\infty}  p_n(x) \hbar^n,\label{powers}
\end{equation}
with \(p_o = p_{cl} =\pm \sqrt{2m(E-V(x))}\). Plugging it in Riccati equation, we require $p_n$'s to obey 
\begin{equation}
	p_n = \frac{i}{2p_o} \left( ip_{n-1}' - \sum_{j = 1}^{n-1}p_i p_{n-i}\right).
\end{equation}
Incorporating these corrections in powers of $\hbar$ in
\begin{equation}
 \Pi_{\gamma}(\hbar) =\oint_{\gamma} p(x) dx = \oint_{\gamma} \sum_n p_n(x) \hbar ^n dx=\sum_{n=0} \Pi_{\gamma,n}(\hbar) \hbar^{n},
 \end{equation}
 and then using the conventional WKB quantization will definitely improve the estimate of the energy spectrum. These $\Pi_{\gamma}(\hbar)$ are known as  {\it quantum periods}  whose classical limit gives  $\Pi_{\gamma,0}(\hbar)$. For the simplest case of QHO, we get corrections from the first order term $\Pi_{\gamma,1}$ leading to the  exact spectrum. Using mathematica, we verified  $\Pi_{\gamma,n>1}$'s  for $n=2,3,4$ are indeed zero. Hence the series (\ref{powers}) converges for QHO. However, it is not clear whether  the series(\ref{powers}) converges for other potentials. In fact, the numerical analysis for monic potentials $V(x)=x^{2M}$ \cite{bender1977numerological} showed
 \begin{equation}
	\Pi_{\gamma,n}(\hbar) = E ^{\frac{1}{2M} + \frac{1}{2} -  n (1 + \frac{2}{2M}) } \frac{2\sqrt{\pi}\Gamma (1 + \frac{1-2n}{2M}) P_n(2M)(-1)^n   }{ \Gamma (\frac{3-2n}{2} + \frac{1-2n}{2M}) (2n+2)! 2^n }. \label{pert}
\end{equation} 
Here, \(P_n\) is a polynomial in \(n\). For \(M= 1\) corresponding to QHO, we see that  $\Pi_{\gamma,n>1}(\hbar)=0$ as
\(\Gamma[ (3-2n))/ 2  + (1-2n)/2 ] \) is  infinite for \(n \geq 2\) confirming only first order correction is non-zero.
However, for  the pure quartic oscillator, the periods $\Pi_{\gamma,n}(\hbar)$ grow without bound for large \(n\). The leading order term in \(P_n(2M)\) is given by 
	$(2n+1)(n+1)(n-1)! B_{2n} (2M)^{ 2n-1} $
where \(B_{2n}\) is the \(2n^{th}\) Bernoulli number. \(P_n(2M)\) grows like $(2n)!$.
This indicates that the quantum period $\Pi_{\gamma}$ can be a divergent series for some potentials. 

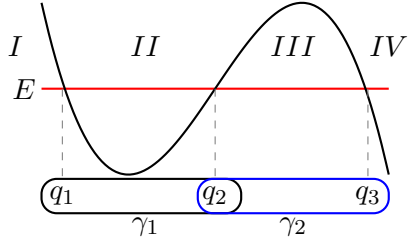
\begin{figure}[ht]
 	\centering
 \begin{tikzpicture}
  \begin{axis}[
 		xmin = -2, xmax = 4,
 		ymin = -5, ymax = 5,
 	axis line style={draw=none},
 	tick style={draw=none},
 	yticklabels={,,},
 	xticklabels={,,}
 	]
 	\addplot[
 			domain = -1:3,
 			samples = 50,
 			smooth,
 			thick,
 			red,
 	] {0};
 	\addplot[
 			domain = -1:3,
 			samples = 50,
 			smooth,
 			thick,
 			black,
 		] {3*x^2-x^3-2};
 	\node at (axis cs:-1.2,0) {\(E\)};
 	\node at (axis cs:-0.76,-2.5) {\(q_1\)};
 	\node at (axis cs:-1.3,1) {\(I\)};
 	\node at (axis cs:0.2,1) {\(II\)};
 	\node at (axis cs:1.9,1) {\(III\)};
 	\node at (axis cs:3,1) {\(IV\)};
 	\node at (axis cs:1,-2.5) {\(q_2\)};
 	\node at (axis cs:2.76,-2.5) {\(q_3\)};
 	\node at (axis cs:0.2,-3.2) {\(\gamma_1\)};
 	\node at (axis cs:1.9,-3.2) {\(\gamma_2\)};
 	\draw[gray] (axis cs:-0.76, 0) -- (axis cs:-0.76, -2.1) [dashed];
 	\draw[gray] (axis cs:1, 0) -- (axis cs:1, -2.1) [dashed];
 	\draw[gray] (axis cs:2.76, 0) -- (axis cs:2.76, -2.1) [dashed];
 	\draw[rounded corners=0.2cm, thick] (axis cs:-1, -2.9) rectangle (axis cs:1.3, -2.1) {};
 	\draw[rounded corners=0.2cm, thick, blue] (axis cs:0.8, -2.9) rectangle (axis cs:3, -2.1) {};
 \end{axis}
 \end{tikzpicture}
 \caption{WKB loops for potential \(V = 3x^2 - x^3\)}
 \label{fig: WKBperiods}
 \end{figure}

In any quantum system, the space of classical configurations is given by the extrema of the potential $V(x)$ (also called saddle points). QHO has one extremum whereas  the cubic potential $V(x) = 3x^2-x^3$ illustrated in Figure \ref{fig: WKBperiods} has two extrema. Technically, we have to investigate 
perturbative series around  each of these classical configurations. They will give different quantum periods $\Pi_{\gamma_1},\Pi_{\gamma_2},\ldots$ For instance, there are two periods in Figure \ref{fig: WKBperiods} corresponding to the curves $\gamma_1$(classically allowed region)  and $\gamma_2$(classically forbidden region). Hence, the divergent series $\Pi_{\gamma}$(\ref {pert}) implicitly signals  the presence of other  perturbative series in the quantum system. This is the theme of resurgent quantum mechanics \cite{delapham}.  In order to capture such information, we need the tools of  `Exact WKB methods'  advocated by Voros for higher degree polynomials \cite{vorosreturn}.   
We will now briefly present the  salient features of the  `Exact WKB method'.

\subsection{Exact WKB Method}

Suppose we split $p(x)=P(x) + Q(x)$(\ref{powers}) involving even and odd powers:
\begin{equation}
	P(x) = \sum_{n=0}^{\infty}p_{2n} \hbar^{2n} ~; ~
	Q(x) = \sum_{n=0}^{\infty}p_{2n+1} \hbar^{2n+1}.
	\label{eq:evenricatti }
\end{equation}
Solving the Riccati equation, we can see odd terms are not independent:
\begin{equation}
	Q(x) = \frac{i\hbar}{2} \frac{d \log(P)}{dx}.
\end{equation}
Hence the general wave function in the classically allowed region can be written as 
\begin{equation} 
\psi(x) = \frac {1}  {\sqrt{P(x)}}  \left(A e^{i\int P(x) dx}+B e^{-i\int P(x) dx} \right)
\end{equation} 
where $A,B$ are normalisation constants . We  need to change  $P(x)$ to $\tilde P(x) = i P(x)$ when we move to the  classically forbidden region.

 In the exact-WKB approach,  the divergent power series will be converted  to a series with finite radius of convergence by a Borel transform.  For  the perturbative series discussed for monic potentials, Borel transform is as follows:
 \begin{equation}
 \Pi_{\gamma}(\hbar) \rightarrow \hat  \Pi_{\gamma}(\xi)= \sum_n \hat \Pi_{\gamma,2n} {\xi}^{2n} = \sum_n \frac{\Pi_{\gamma,2n}}{2n!} {\xi}^{2n},
 \end{equation} 
 which is analytic near origin in the complex plane $\xi$. Then, the $ \hat  \Pi_{\gamma}(\xi)$ is promoted to a function through a procedure called  `Borel resummation' :
\begin{equation}
 B_{\phi}[\Pi_{\gamma}]( \hbar)= \frac{1}{  \hbar} \int_{0}^{e^{i\phi}\infty} e^{-\xi /\hbar}  \hat  \Pi_{\gamma}(\xi) d \xi~, ~~\hbar \in \mathbb R_{>0}.
 \end{equation}
 The above  integral denotes a Laplace transform  of $\hat  \Pi_{\gamma}(\xi)$ along a direction, defined by angle $\phi$, in the complex plane $\xi$. If the integral converges for small $\hbar$, then the corresponding  quantum period $\Pi_{\gamma}(\hbar)$ is said to be Borel summable. Suppose $\hat  \Pi_{\gamma}(\xi)$  has singularities on the complex plane $\xi$, then the Borel summability cannot be performed on the rays containing such  singularities. For instance, a simple pole at \(\xi_0\) whose $\arg\xi_0= {\chi}$ will imply a discontinuity in the Borel resummation. That is., 
 $${\rm \lim}_{\delta \rightarrow 0}~ \mathcal{B}_{\chi+\delta}[\Pi_{\gamma}](\hbar) \neq ~{\rm \lim}_{\delta \rightarrow 0} ~\mathcal{B}_{\chi-\delta}[\Pi_{\gamma}]( \hbar).$$
 Hence, we  define median Borel resummation $\mathcal{B}_{\chi}^{med}[\Pi_{\gamma}]( \hbar)$, the lateral Borel resummation \(\mathcal{B}_{\chi_{\pm}}[\Pi_{\gamma}](\hbar)\) and the 
 Stokes discontinuity  ${\rm disc}_{\chi}  [\Pi_{\gamma}]( \hbar)$ to characterise and overcome such obstructions to Borel summability:
 \begin{eqnarray}
 \mathcal{B}_{\chi}^{med}[\Pi_{\gamma}]( \hbar)&=&{\frac{1}{2}\rm \lim}_{\delta \rightarrow 0}~ \mathcal{B}_{\chi+\delta}[\Pi_{\gamma}](\hbar)+\mathcal{B}_{\chi-\delta}[\Pi_{\gamma}] (\hbar)\\
 B_{\chi_{\pm}}[\Pi_{\gamma}](\hbar) &=&  {\rm \lim}_{\delta \rightarrow 0}~ B_{\chi\pm \delta}[\Pi_{\gamma}](\hbar) \nonumber \\ 
 {\rm disc}_{\chi}[\Pi_{\gamma}]( \hbar)&=&{\rm \lim}_{\delta \rightarrow 0}~ B_{\chi+\delta}[\Pi_{\gamma}]( \hbar)-B_{\chi-\delta}[\Pi_{\gamma}](\hbar).\nonumber\label{discontinuty}
 \end{eqnarray}
 The knowledge of all the Stokes discontinuities as well as  the classical limit  $\Pi_{\gamma,0}$ of the quantum periods are  required to reconstruct the quantum periods (as solutions to the Riemann-Hilbert problem).  The {\it  Delabaere-Pham formula } \cite{delapham,iwaki2014exact} encodes the  structure of discontinuities of  any quantum period in terms of the other quantum periods:
	\begin{equation}
  	\mathcal{B}_{\chi -}(\mathcal{V}_{\gamma_i}) = \mathcal{B}_{\chi +}(\mathcal{V}_{\gamma_i}) \prod_{j \neq i} (1 + \mathcal{V}_{\gamma_j}^{-1} )^{-(\gamma_i,		\gamma_j)}. 
	\label{eq:DelabaerePhamgeneral}
	\end{equation}
Here,
$$\mathcal{V}_{\gamma_i} = \exp{( \frac{ i\Pi_{\gamma_i}} {\hbar})}$$ is called as {\it Voros symbol} and \( (\gamma_i, \gamma_j) \) is the intersection number between the curves \( \gamma_i, \gamma_j \). Once we have the solutions for all the Voros symbols, 
the {\it exact WKB connection formula} (also known as Voros-Silverstone connection formulae) leads to an {\it exact quantization condition} (EQC) as a single functional relation between $\mathcal{V}_{\gamma_i}$'s:
\begin{equation}
	f(\mathcal{V}_{\gamma_1}, \mathcal{V}_{\gamma_2},\dots  ) = 0.\label{eqc}
\end{equation}
For example,  using the Voros-Silverstone connection formulae for the cubic potential $V(x) = 3x^2-x^3$, the following EQC relating the two quantum periods (as drawn in Figure \ref{fig: WKBperiods}) can be deduced \cite{ito2019tba,emery2021tba}:
\begin{equation}
	2 \cos \left(\frac{1}{2\hbar } \mathcal{B}_{\chi \pm}(\Pi_{\gamma_1})  \right) + \exp \left(-\frac{i}{\hbar} \Pi_{\gamma_2} \right) = 0
\end{equation}
Recall that the $\Pi_{\gamma_2}$ is associated with the classically forbidden region. Such a relation gives the  values of energy \(E_n\). Sometimes it is convenient to  fix the value of energy and compute the values of \(\hbar_n(E)\) for which the EQC(\ref {eqc}) holds. These values of \(\hbar_n(E)\) are called  {\it Voros spectrum}. 

As mentioned earlier, the solution to the Riemann-Hilbert problem (quantum periods) can be obtained from a set of `{\it Thermodynamic Bethe Ansatz}'(TBA)  equations. We will briefly discuss TBA method in the following section. 
 \section{TBA system \label{gensec:TBA}}
 For monic potentials, including quartic oscillator, $V(x)=x^{2M}$, there are $2M$ turning points located at $\{w^i E^\frac{1} { 2M}\}$
in the complex plane  where  $\omega$ is $2M$-th root of unity and $E$ is the energy. Only two of the turning points are on the real axis.  Similarly, for a general $d$-degree polynomial potentials $V(x) = \sum_{n=1}^d a_n x^n$, there will $d$ turning points. Depending on the choice of $a_n$ (known as moduli), some of the turning points could be real or complex. We can make a suitable choice of the moduli so that all the turning points $x_1 < x_2 < \ldots < x_d$ are on the real axis. This choice is sometimes referred to as `{\it minimal chamber}' in the literature. Such a minimal chamber will allow $\lfloor(d-1)/2\rfloor$ cycles $\{\gamma_i\}$. In fact, the  cubic potential $V(x)=3x^2-x^3$  shown in Figure \ref{fig: WKBperiods}  allows two cycles $\gamma_1, \gamma_2$ in the minimal chamber.

 In such a minimal chamber,  the quantum periods $\Pi_{\gamma_{2i}}$ corresponding to classically forbidden region are Borel summable along the positive real axis of $\hbar$ whereas $\Pi_{\gamma_{2i-1}}$ corresponding to classically allowed region is not Borel summable. Hence the  discontinuity formula(\ref{eq:DelabaerePhamgeneral}) along the real line ($\chi=0$) is 
 \begin{equation}
 {\rm disc}_{0} \Pi_{2i-1}= -i \hbar  \log(1+ \mathcal V_{2i-2}^{-1})-i \hbar  \log(1+ \mathcal V_{2i}^{-1}).
 \end{equation}
 Similarly, there is a discontinuity at $\chi=\pi/2$ for the quantum periods $\Pi_{2i}$ whereas $\Pi_{\gamma_{2i-1}}$ are Borel summable. These two situations are neatly incorporated by defining \(\epsilon_a\) functions as:
\begin{align}
	-i\epsilon_{2i-1} (\theta + i\pi/2 \pm i\delta) &= \frac{1}{\hbar} \mathcal{B}_{0 \pm} (\Pi_{\gamma_{2i-1}})(\hbar) \\
	-i\epsilon_{2i} (\theta) &= \frac{1}{\hbar} \mathcal{B} (\Pi_{\gamma_{2i}}),
\end{align}
where $e^{\theta} = 1/\hbar$ \cite{ito2019tba}.
Clearly, these $\epsilon_a$ functions have a discontinuity at  \(\chi =\pi /2\) for both even and odd $a$. Hence, the Delabaere-Pham discontinuities(\ref{eq:DelabaerePhamgeneral}) can be compactly written as:
\begin{equation} 
{\rm disc}_{\pi /2} \epsilon_a(\theta) = L_{a-1} (\theta) + L_{a+1}(\theta)~, {\rm where}~ L_a=\log(1+ e^{-\epsilon_a(\theta)}).\label{discepsi}
\end{equation}
Further, the asymptotic series  of the functions $\epsilon_a(\theta)$ will be
\begin{equation} 
\epsilon_i(\theta) = m_a e^{\theta} + \mathcal O(e^{-\theta}),
\end{equation}
 where $m_a$'s, referred to as masses in two-dimensional integrable theories, are the classical periods:
 \begin{equation}
 m_a = \Pi_{\gamma_a,0}= \oint_{\gamma_a} P(x) dx= 2 \int_{x_a}^{x_{a+1}} P(x) dx~{\rm where}~\gamma_a=[x_a,x_{a+1}].\label{asymmass}
 \end{equation}
Remember to replace $P(x) \rightarrow i P(x)$ whenever the cycle $\gamma_{a \equiv 2i}$ ( classically forbidden region) so that $m_a$'s are real and positive.
 
 The solution to  the Riemann Hilbert problem for the functions $\epsilon_a(\theta)$ obeying (\ref {discepsi}) and (\ref{asymmass})  can be obtained  using the following system  of TBA integral equations in the minimal chamber:
\begin{equation}
	\epsilon_a(\theta) = m_a e^{\theta} - \int_{\mathbb{R}} \frac{L_{a-1} (\theta')} {\cosh(\theta - \theta')} d\theta' - \int_{\mathbb{R}} \frac{L_{a+1} (\theta')} {\cosh(\theta - \theta')} d\theta' ~a=1,2,\ldots d-1
	\label{eq:TBAeqs}
\end{equation}
As $P(x)$ is a series in even powers of $\hbar$, we have to take both $\hbar$ positive as well as negative. This in turn adds another similar discontinuity equation, and combining all of these discontinuities transforms the usual propagator into the \(\sinh\) propagator. Finally, the rotation by 
$\pi/2$ gives us the $\cosh(\theta- \theta')$ in the integral equation\footnote{We thank Katsushi Ito for clarifying this point.}.
For other choices of moduli$\{a_n\}$ in the potential $V(x)=\sum_n a_n x^n$, some  turning points can be on the complex plane. This leads to additional periods in the complex plane. We do not review calculations involving complex turning points here. This is discussed in great detail in \cite{emery2021tba}.

\section{TBA for \(|x|\)  \label{sec:TBAmodx}}

We have seen in the previous section how the TBA system can be used to compute  quantum periods for the polynomial potentials which are smooth \cite{DoreyTateo2007,ito2019tba} and deduce the Voros spectrum from EQC.  Interestingly, the TBA equations have been extended to the potentials with regular singularities like \(1/x^2\) \cite{ito2020tba}. 
However, the EQC for such singular potentials has not been attempted. Conversely for the $|x|$ potential, we know the true spectrum but not the TBA equations incorporating the derivative singularity at the origin.

Note that the potentials of the form \(|x|^n\) with  \(n\)  odd positive integer were considered in \cite{voros1999airy} using exact WKB method and spectral determinants but without TBA equations.
A TBA equation was also derived in \cite{fendley1999airy} albeit from very different considerations of  \(\mathcal{N} = 2 \) supersymmetric field theories. In fact, this  was argued to be the spectral determinant of the \(|x|\) potential which was expanded upon in \cite{suzuki2015elementary}. These developments do suggest that there could be a TBA approach for the \(|x|\) potential. In the following subsection, we discuss the naive TBA approach for the  \(|x|\) potential and show that the energy spectrum do not match the true spectrum for low lying states.

  \subsection{Naive TBA approach for \(|x|\)}\label{naivetba}
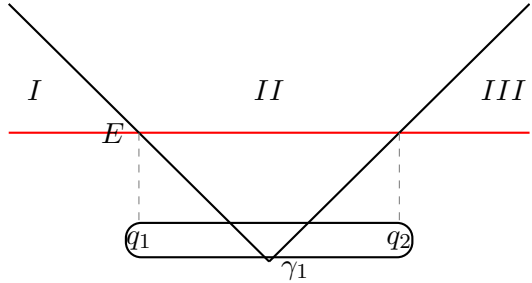
\begin{figure}[ht]
	  \centering
  \begin{tikzpicture}
   \begin{axis}[
		  xmin = -2, xmax = 2,
		  ymin = -5, ymax = 5,
	  axis line style={draw=none},
	  tick style={draw=none},
	  yticklabels={,,},
	  xticklabels={,,}
	  ]
	  \addplot[
			  domain = -2:2,
			  samples = 50,
			  smooth,
			  thick,
			  red,
	  ] {0};
	  \addplot[
			  domain = 0:2,
			  samples = 50,
			  smooth,
			  thick,
			  black,
		  ] {3*x-3};
		  \addplot[
			  domain = -2:0,
			  samples = 50,
			  smooth,
			  thick,
			  black,
		  ] {-3*x-3};
	  \node at (axis cs:-1.2,0) {\(E\)};
	  \node at (axis cs:-1.8,1) {\(I\)};
	  \node at (axis cs:0,1) {\(II\)};
	  \node at (axis cs:1.8,1) {\(III\)};
	  \node at (axis cs:-1,-2.5) {\(q_1\)};
	  \node at (axis cs:1,-2.5) {\(q_2\)};
	  \node at (axis cs:0.2,-3.2) {\(\gamma_1\)};
	  \draw[gray] (axis cs:-1, 0) -- (axis cs:-1, -2.1) [dashed];
	  \draw[gray] (axis cs:1, 0) -- (axis cs:1, -2.1) [dashed];
	  \draw[rounded corners=0.2cm, thick] (axis cs:-1.1, -2.9) rectangle (axis cs:1.1, -2.1) {};
  \end{axis}
  \end{tikzpicture}
  \caption{WKB loops for potential \(V = 3|x| \)}
  \label{fig: WKBperiodnaivemodx}
  \end{figure}

  Let us  blindly apply the  TBA tools, applicable for polynomial potentials, to deduce the WKB periods for the $|x|$ potential. The Schr\"{o}dinger equation is given by  
  \begin{equation}
	  \hat{H}\psi = -\frac{\partial ^2}{\partial x^2 } \psi + |x|\psi= E\psi  .
   \end{equation}
 In this case, there are two turning points, corresponding to \(x = +E\) and \(x  = -E\).  As there is only one nontrivial cycle \(\gamma\) between them as seen in Figure \ref{fig: WKBperiodnaivemodx}, the solution to the TBA equation for this period is the mass $m$. As there is only one cycle, we believe that the EQC for $|x|$ will be similar to the QHO case:
  \begin{equation}
	  \frac{\Pi_{\gamma}}{\hbar} = \frac{m}{\hbar}  = 2\pi \left(n + \frac{1}{2}\right), ~~ n=0,1,2\dots
	  \label{eq:modxnaivetba}
  \end{equation} 
  where the \(0^{th}\) order mass \(m\) is given by 
  \begin{equation}
	  m = \oint_{\gamma} \sqrt{E - |x|}~dx = 2 \int_{-E}^E \sqrt{E - |x|}~dx = \frac{8}{3}\frac{E^{\frac{3}{2}}}{ \hbar}
  \end{equation}
  On solving the Bohr-Sommerfeld quantisation condition(\ref{eq:modxnaivetba}) we obtain the spectrum for \(\hbar = 1, 2m = 1\)
	  \begin{equation}
		 E_n = \left(\frac{3\pi}{4}\left(n+\frac{1}{2} \right) \right)^{(2/3)}.
	  \end{equation}
 Table \ref{fig:datatableAiry} shows that the true spectrum \cite{englert2020particle} matches well with the conventional WKB results for \(n\geq 5\). Such a mismatch at low  \(n\) implies that the naive analogy of EQC between $|x|$ and QHO  is not correct.

	  \begin{table}
		  \center{
		 \begin{tabular} {c|c|c}
		 $n$ & True Spectrum & Naive TBA Spectrum \\ \hline
		 0 & 1.01879 & 1.1154602372253557 \\ 
		 1 & 2.33811 & 2.320250794710102 \\ 
		 2 & 3.2482 & 3.2616255199180713 \\ 
		 3 & 4.08795 & 4.081810015382323 \\ 
		 4 & 4.8201 & 4.826316143499807 \\ 
		  5 & 5.52056 & 5.517163872783549 \\ 
		  6 & 6.16311 & 6.167128465231806 \\ 
		  7 & 6.78311 & 6.784454480834836 \\ 
		  8 & 7.3721 & 7.374853108941933 \\ 
		  9 & 7.94413 & 7.942486663292496 \\ 
		 
		  \end{tabular}}
		  \caption{Spectrum for \(|x|\) potential \(n=0\) to \(9\). }
		  \label{fig:datatableAiry}
	  \end{table}
	  
This exercise clearly indicates that we must introduce some modifications to incorporate the derivative singularity at $x=0$. Taking clue from our modified Bethe-like ansatz,  we will show that the  generalisation of the TBA equations to \(|x|\) is indeed possible by reinterpreting the problem on a half real line similar to \cite{voros1999airy}. We will now discuss the subtle features in the following subsection.

	  \subsection{Effective radial problem for \(|x|\)  }
The most natural way to consider the \(|x|\) problem is to treat $x$ as radial coordinate $r$. The Schr\"{o}dinger equation in radial coordinate is 
	  \begin{equation}
		   \hat{H}\psi = \left(\frac{-\hbar^2}{2m} \frac{\partial^2}{\partial r ^2} + r \right)\psi= E\psi .
     \label{effrad}
	  \end{equation} 
   
Recall that the radial part of the differential equation describing hydrogen atom contains a centrifugal term \( l (l + 1) / r^2 \) 
($l$ denotes orbital angular momentum). We believe that the derivative singularity can be made to appear as a centrifugal term. This will lead to a correction to the potential in the $r$ coordinate.  

By performing the change of variable $x \rightarrow r$  the derivative discontinuity has seemingly vanished. However, the singularity is still present in the topology of the problem. That is,  \(\psi \) is now a function of  \(r \in  [0 , \infty)\). Hence we need to specify a boundary value for \(\psi\) at \(r=0\) instead of the exponential decay as  \(x \rightarrow -\infty\). 

Note that the parity symmetry of $|x|$ potential requires that the wave function must be either symmetric or anti-symmetric: $\psi(x) = \pm \psi(-x)$. In the $r$-coordinate system, such a parity symmetry imposes either Dirichlet ($\psi(r=0)=0$) or Neumann boundary conditions  ($\psi^{\prime}(r)\vert_{r=0}=0$).
Consider the wave function near $r=0$ to be of the following form:
\begin{equation}
		  \psi(r) \sim r^{\ell} f(r),   
	  \end{equation}
where \(f(r)\) is some function nonzero at the origin whose derivative \(f^{\prime}(r) \vert_{r=0} \) vanishes. The allowed boundary value conditions 
	  \begin{align*}
		  \psi(0) =  0 ,~\psi'(0) \neq 0 ~~;~~
		  \psi(0) \neq 0,~ \psi'(0) = 0, 
	  \end{align*}
forces  \(\ell \) to be either 1 or 0 respectively. 
Therefore, for the $|x|$ potential in radial coordinate, we can add the following centrifugal term as
	  \begin{equation}
		  \frac{\hbar^2 (\ell)(\ell-1) }{r^2}, \label{eq:elltol}
	  \end{equation}  
 to do the TBA calculations. In other words, the singularity at $x=0$ is traded for an effective centrifugal term. This corrected potential is a subset of a more general  potential with single and double pole \cite{ito2020tba} for a suitable choice of the parameters.

 We will now present the salient features of the TBA system for potential with single and double pole \cite{ito2020tba}. This sets the stage to numerically compute the quantum periods for the linear potential $|x|$ by taking suitable limits for the parameters.

	    \subsection{TBA equation for a potential with Single and Double Pole \label{sec:TBApole}}
	  We will briefly review the Schr\"{o}dinger type equation with polynomial potentials 
   with simple pole  and a centrifugal term \cite{ito2020tba}:
   \begin{equation}
    \Bigl(-\hbar^2 \frac{d^2}{dx^2} + x^{s+1}+\sum_{a=1}^{s+2} u_a x^{s+1-a}+\hbar^2 \frac{l(l+1)}{x^2}\Bigr)\psi(x)=0.
    \label{eq:SEtypewithpole}
\end{equation}
Here $x \geq 0, s\geq 0$, $l$ is any real number and $u_a$'s are
parameters.
Note that $\ell = l+1$(\ref{eq:elltol}). For the following choice of parameters:
   \begin{equation}
  s=0~;~ u_1=-E ,
   \end{equation} 
  the (\ref{eq:SEtypewithpole}) reduces to 
	  \begin{equation}
		  \Bigl(-\hbar^2 \frac{d^2}{dx^2} + \frac{x^2-Ex+u_2}{x}+\hbar^2 \frac{l(l+1)}{x^2}\Bigr)\psi(x)=0 .\label{linearpotapprox}
	  \end{equation}
 This resembles the equation for the linear potential (\ref{effrad}), in the limit  $u_2 \rightarrow 0, l \rightarrow 0 ~{\rm or}~-1$. We will present the TBA equation for the potential (\ref{linearpotapprox}) as discussed in \cite{ito2020tba}. 
   
\begin{figure}[ht]
		  \centering
	  \begin{tikzpicture}
	   \begin{axis}[
			  xmin = -1, xmax = 4,
			  ymin = 0, ymax = 5,
		  axis line style={draw=none},
		  tick style={draw=none},
		  yticklabels={,,},
		  xticklabels={,,}
		  ]
		  \addplot[
				  domain = -1:3,
				  samples = 50,
				  smooth,
				  thick,
				  black,
		  ] {1};
		  \addplot[
				  domain = 0.01:3,
				  samples = 800,
				  smooth,
				  thick,
				  black,
			  ] { x + (1/x) };
			  \addplot[
				  domain = -1:3,
				  samples = 200,
				  smooth,
				  thick,
				  black,
			  ] { 3 };
		  \node at (axis cs:-1.2,1) {\(E\)};
		  \node at (axis cs:-0.1,0.8) {\(0\)};
		  \node at (axis cs:0.38,0.8) {\(e_1\)};
		  \node at (axis cs:2.6,0.8) {\(e_2\)};
		  \node at (axis cs:0.2,2.5) {\(\hat{\gamma}\)};
		  \node at (axis cs:1.5,2.5) {\(\gamma_1\)};
		  \draw[gray] (axis cs:0, 1) -- (axis cs:0, 3) [dashed];
		  \draw  (axis cs:0,0 )-- (axis cs:0,5) [thick];
		  \draw[gray] (axis cs:0.38, 1) -- (axis cs:0.38, 3) [dashed];
		  \draw[gray] (axis cs:2.6, 1) -- (axis cs:2.6, 3) [dashed];
		  \draw[rounded corners=0.1cm, thick] (axis cs:-0.1, 3.2) rectangle (axis cs:0.4, 2.8) {};
		  \draw[rounded corners=0.2cm, thick, blue] (axis cs:0.37, 3.2) rectangle (axis cs:2.7, 2.8) {};
	  \end{axis}
	  \end{tikzpicture}
	  \caption{WKB loops for potential \(V = x  + u_2/x \)}
	  \label{fig:WKBperiodsmodx}
	  \end{figure}
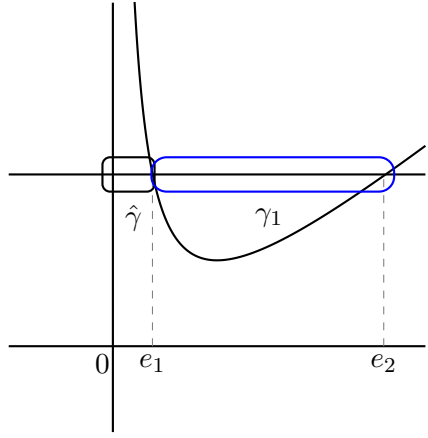

Taking  $x \in \mathbb{C}$ (complex  domain), the equation (\ref{linearpotapprox}) remains invariant under Symanzik rotation: 
   \begin{equation}
    (x,E,u_2)\to (\omega x, \omega E, \omega^2 u_2) ~,~
    {\rm where}~ \omega=\exp{\frac{2\pi i}{s+3}}\vert_{s=0}.
    \end{equation}
From the semi classical behaviour in the limit \(\hbar \to 0\), the
turning points $e_1,e_2$ for 
$$E=V(x)=\frac{x^2+u_2}{x},$$
can be chosen to be in the positive real axis: $0<e_1<e_2$ for $u_2 \geq 0$. Note that $e_1 \rightarrow 0$ as $u_2 \rightarrow 0$. For this potential, there are two cycles as shown in 
Figure \ref{fig:WKBperiodsmodx}: $\gamma_1$ encircling $e_1$ and $e_2$ (classically allowed region) and $\hat{\gamma}$ encircling the  pole at the origin $0$ and $e_1$ (classically forbidden).
The $l$ dependent centrifugal term (double pole) is responsible for a non-trivial monodromy of the wave function around the origin. This is very similar to  our modified Bethe-like ansatz in section \ref{sec:HABethe} for the hydrogen atom potential. In fact, the non-trivial monodromy and the Symanzik rotation symmetry leads to modification of the TBA equations\footnote{Interested readers can see \cite{ito2020tba} for details} for the $\gamma_1$ and $\hat {\gamma}$ cycles:
	  \begin{eqnarray}
	  \epsilon_1(\theta)&=& m_1e^\theta - \int_{\mathbb{R}} \frac{\log(1+\omega^{3/2}e^{2\pi il}e^{-\hat{\epsilon}(\theta')})+\log(1+\omega^{3/2}e^{-2\pi il}e^{-\hat{\epsilon}(\theta')})} {\cosh(\theta - \theta')} \frac{d\theta'}{2\pi}  \nonumber \\ 
		\hat{\epsilon}(\theta)&=& \hat{m}e^\theta-  \int_{\mathbb{R}}\frac{\log(1+e^{-\epsilon_1(\theta')})}{\cosh(\theta-\theta')}\frac{d\theta'}{2\pi}.
		\label{eq:TBAsingledoublepole}
	  \end{eqnarray}
 Here $m_1,\hat{m}$ are the $0^{th}$ order WKB periods chosen with orientation so that they
	  are both real and positive:
	  \begin{equation}
		  m_1 =\frac{1}{i} \oint_{\gamma_1} \sqrt{\frac{x^2 - Ex + u_2}{x}}dx ~~;~~
		  \hat{m} = \oint_{\hat{\gamma}} \sqrt{\frac{x^2 - Ex + u_2 }{x} } dx .
	  \end{equation}
As discussed in section \ref{gensec:TBA}, the classically allowed period $\Pi_{\gamma_1}$ is not Borel summable along the positive real axis of $\hbar$. So, the period is resummed by taking average of two  Borel resummations(lateral and median resummation(\ref{discontinuty})) calculated just after and before crossing the discontinuity along the positive real $\hbar$ axis. For the above TBA equations, the median resummation is \cite{ito2020tba}
	  \begin{equation}
		\frac{1}{\hbar}\mathcal{B}_{med}(\Pi_{\gamma_1})(\hbar)=m_1e^\theta+P\int_{\mathbb{R}}\frac{\log(1+\omega^{3/2}e^{2\pi il}e^{-\hat{\epsilon}(\theta')})(1+\omega^{3/2}e^{-2\pi il}e^{-\hat{\epsilon}(\theta')})}{\sinh(\theta-\theta')}\frac{d\theta'}{2\pi}
	  \end{equation}
	  where P is the principal value of the integral which can be computed using the formula:
	  \begin{equation}
		P\int_{\mathbb{R}}\frac{f(\theta')}{\sinh(\theta-\theta')}d\theta'= \lim_{\delta \to 0}\int_{\mathbb{R}}\frac{\sinh(\theta-\theta')\cos(\delta)}{\sinh^2(\theta-\theta')\cos^2(\delta)+\cosh^2(\theta-\theta')\sin^2(\delta)}f(\theta')d\theta'
	  \end{equation}
   Note that the TBA equations are same for any integer $l$. 

  At this stage it is natural to return to the $|x|$ problem by taking the limit $l \to 0 , u_2 \to 0$. However, this limit runs into several singularities which need regularisation. Some of these details are discussed in Appendix \ref{b.Regularisation of TBA}. Instead we circumvent this problem by keeping $u_2, l$ small but finite in our computation. 
 \subsubsection{Numerical Computation of quantum periods for $|x|$}
To solve the TBA system, we used the Gaussian Interpolation technique presented in Appendix B of \cite{emery2021tba} with \(2^{12}\) points randomly distributed around \(\theta = 0\) instead of the Fourier transform method used in \cite{ito2020tba}. 

\begin{figure}
	\centering
\begin{subfigure}{0.48\textwidth}
	\centering
	\includegraphics[width=\textwidth]{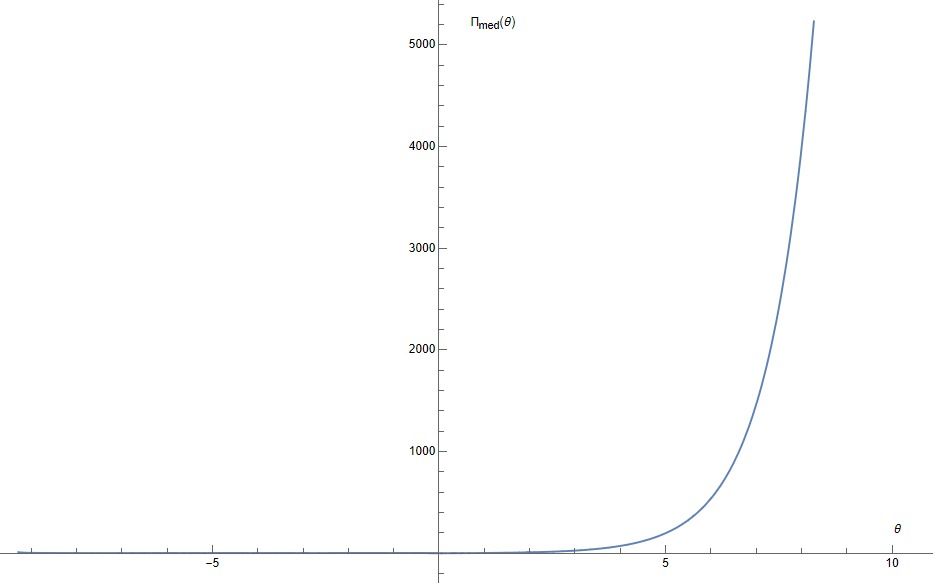}
	\caption{Median resummed period $\mathcal{B}_{med}(\Pi_{\gamma_1})$  }
	\label{fig: medianresummedmodx}
\end{subfigure}
\begin{subfigure}{0.48\textwidth}
	\centering
\includegraphics[width=\textwidth]{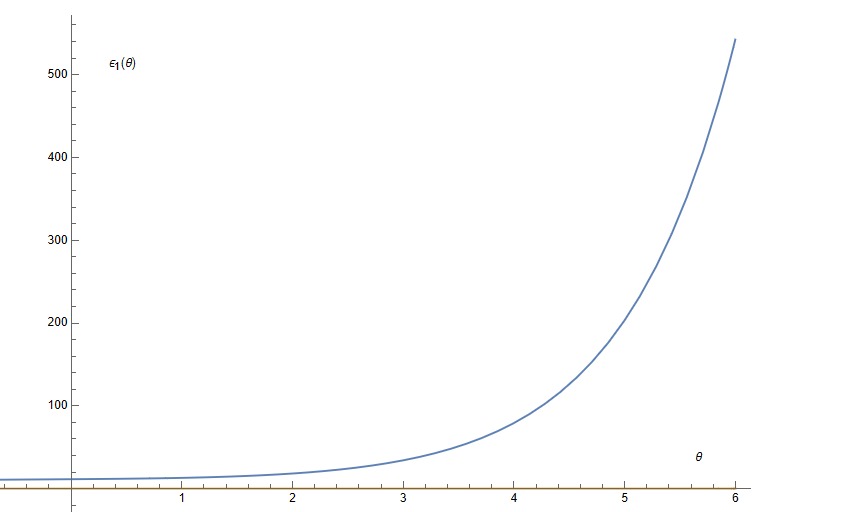}
	\caption{ Solutions to TBA equations of $\epsilon_1$}
	\label{fig: epsilonmmodx1}
\end{subfigure}
\begin{subfigure}{0.48\textwidth}
	\centering
\includegraphics[width=\textwidth]{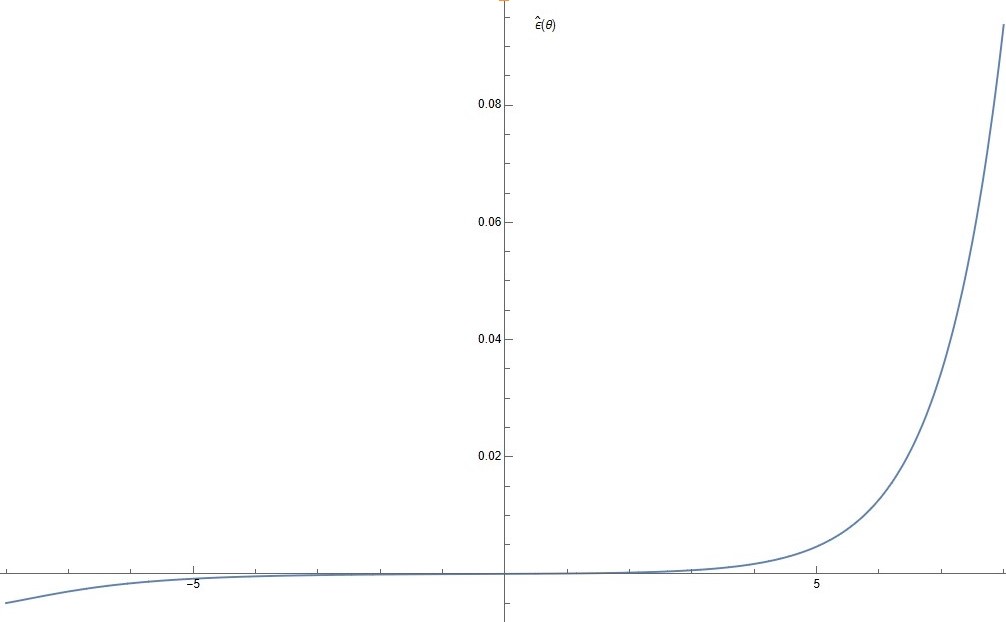}
\caption{ Solutions to TBA equations of $\epsilon_2$}
	\label{fig: epsilonmmodx}
\end{subfigure}
\caption{$ V(x) \sim |x|$  for \(E = 1,u_2=10^{-8},l = 10^{-5}\)}
\label{fig:quantperiod}
\end{figure}

First, we validated our numerical code on quantum periods by confirming that the Voros spectrum 
   for  \(u_1 = -3, u_2 = 1, l = -2/5 \) matched with the Table 2 in \cite{ito2020tba} obtained using Bohr-Sommerfeld quantization:
	  \begin{equation}
		\frac{1}{\hbar}\mathcal{B}_{med}(\Pi_{\gamma_1})(\hbar) \sim  2\pi (k+1/2) ~~~~ k\in \mathbb{Z}_{\ge 0}. \label{eq:incorrecteqc}
	  \end{equation} 
Using our validated numerical code, we obtained the quantum periods for \(E = 1, u_2=10^{-8}, l = 10^{-5}\). These are plotted in Figure \ref{fig:quantperiod}. A mathematica file containing this computation is linked on the arXiv page as an ancillary file. 

Our next step is to obtain Voros spectrum using these quantum periods. This requires deriving exact quantization conditions (EQC) for potentials which are singular at the origin. Bohr Sommerfeld quantization (\ref{eq:incorrecteqc}) is not correct to determine $\theta_n$ for low lying ground and excited states for potentials(\ref{linearpotapprox}). We will now address the EQC for potentials with singular behaviour near the origin.

\section{Exact quantization condition}
\label{sec:EQC}
The key construction to arrive at the TBA equations for the \(|x|\) potential was a shift to radial coordinates \(r =|x|\) with an additional centrifugal term \(\hbar l (l +1) / r^2\) (vanishing for Dirichlet or Neumann boundary conditons at $r=0$). Even though, we obtained the quantum periods $\Pi_{\gamma_1}, \Pi_{\hat {\gamma}}$ solving numerically the TBA system of equations, we have no idea how to write 
Voros-Silverstone connection formula to deduce the exact quantization condition (EQC).

To deal with the pole at origin we look into our modified Bethe-like ansatz proposal for hydrogen atom in section \ref{sec:BetheLike}. There, the pseudomomentum(\ref{eq:bethelike-withl}) with an additional  orbital angular momentum dependent simple pole at the origin reflects the zero of the wave function at the origin.
This suggests that in the EQC we may have to introduce additional correction to the quantum periods enclosing the origin. 
{\it This is not needed for polynomial potential which has no singular behaviour at the origin.}

From our modified Bethe-like ansatz(\ref{eq:bethelike}) as well as the exact solution for the symmetric potential $V(x)=-\frac{1}{ |x|}$ \cite{ishkhanyan2019wkb,ishkhanyan2017}, the quantization condition for $l=0$ is
\begin{equation}
\oint p ~dx = 2 \pi \hbar (N+1)~,~N=0,1\ldots .\label{eq:eqchydro}
\end{equation}
We put forth the following proposal:\\
{\bf Proposition 3:}
\label{prop3}
The correction in the EQC to the quantum period due to singular behaviour at the origin for $l=0$ is
\begin{equation}  
\Pi_0 = 2 \pi \hbar. \label{eq:origincontri}
\end{equation}
Technically this should also be the correction for $l=-1$ as the potential is unchanged.

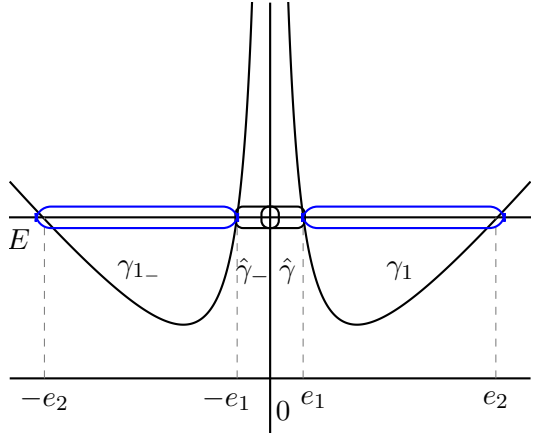
\begin{figure}[ht]
	\centering
\begin{tikzpicture}
 \begin{axis}[
		xmin = -3, xmax = 3,
		ymin = 1, ymax = 5,
	axis line style={draw=none},
	tick style={draw=none},
	yticklabels={,,},
	xticklabels={,,}
	]
	\addplot[
			domain = -3:3,
			samples = 50,
			smooth,
			thick,
			black,
	] {1.5};
	\addplot[
			domain = 0.1:3,
			samples = 800,
			smooth,
			thick,
			black,
		] { x + (1/x) };
         \addplot[
			domain = -3:-0.01,
			samples = 800,
			smooth,
			thick,
			black,
		] {- x - (1/x) };
		\addplot[
			domain = -3.2:3.2,
			samples = 200,
			smooth,
			thick,
			black,
		] { 3 };
	\node at (axis cs:-2.9,2.8) {\(E\)};
        \node at (axis cs:0.15,1.2) {\(0\)};
        \node at (axis cs:0.5,1.3) {\(e_1\)};
        \node at (axis cs:-0.5,1.3) {\(-e_1\)};	
        \node at (axis cs:2.6,1.3) {\(e_2\)};
	\node at (axis cs:0.2,2.5) {\(\hat{\gamma}\)};
	\node at (axis cs:1.5,2.5) {\(\gamma_1\)};
	\node at (axis cs:-2.6,1.3) {\(-e_2\)};
	\node at (axis cs:-0.2,2.5) {\(\hat{\gamma}_-\)};
	\node at (axis cs:-1.5,2.5) {\(\gamma_{1_-}\)};
	
	\draw  (axis cs:0,0 )-- (axis cs:0,5) [thick];
	\draw[gray] (axis cs:0.38, 1.5) -- (axis cs:0.38, 3) [dashed];
	\draw[gray] (axis cs:2.6, 1.5) -- (axis cs:2.6, 3) [dashed];
        \draw[gray] (axis cs:-0.38, 1.5) -- (axis cs:-0.38, 3) [dashed];
	\draw[gray] (axis cs:-2.6, 1.5) -- (axis cs:-2.6, 3) [dashed];
	\draw[rounded corners=0.1cm, thick] (axis cs:-0.1, 3.1) rectangle (axis cs:0.4, 2.9) {};
	\draw[rounded corners=0.2cm, thick, blue] (axis cs:0.37, 3.1) rectangle (axis cs:2.7, 2.9) {};
        \draw[rounded corners=0.1cm, thick] (axis cs:0.1, 3.1) rectangle (axis cs:-0.4, 2.9) {};
        \draw[rounded corners=0.2cm, thick, blue] (axis cs:-0.37, 3.1) rectangle (axis cs:-2.7, 2.9) {};
	\end{axis}
\end{tikzpicture}
\caption{WKB loops for potential \(V = |x|  + \frac{1} {|x|} \)}
\label{fig: WKBperiodsmodxEQC1}
\end{figure}

In order to work out EQC, we would like to go back to
$x \in (-\infty, \infty)$ where the 
wave function decays as $x \rightarrow \pm \infty$. Clearly, the radial TBA system is symmetrically mirrored about the origin. For the \(|x|\) case in $x$ domain, there are 2 more loops, as shown in Figure \ref{fig: WKBperiodsmodxEQC1} which we will denote by \( \gamma_{1_-} ,\hat{\gamma}_-\). From the symmetry $V(x)=V(-x)$,  we expect 
\begin{align*}
	\Pi_{\gamma_1} = \Pi_{\gamma_{1_-}} , ~~~\Pi_{\hat{\gamma}} = \Pi_{\hat{\gamma}_- } .
\end{align*}
However, the TBA system should continue to be exactly the same for \(\Pi_{\gamma_1},\Pi_{\hat{\gamma}}\) and their negative analogue. In fact, this equivalence is mainly due to the fact that the {\it origin is not a turning point} and hence should not contribute any additional Borel resummation discontinuities in the periods containing it. We see that from the period structure, the TBA system for \(|x|\) is analogous to that of the TBA system for the symmetric double well potential, except for the singular behaviour at the origin.
Hence we propose the following for the classically forbidden cycle $\hat {\Gamma}$ between the two turning points $\pm e_1$:

\noindent
{\bf Proposition 4:}\\
 The quantum period for the  classically forbidden cycle $\hat {\Gamma}$
between the turning points $\pm e_1$  will be 
\begin{equation}
	\Pi_{\hat{\Gamma}} =\Pi_{\hat{\gamma}} +  \Pi_{\hat{\gamma}_- } + i\Pi_0, \label{eq:closedcycle}
\end{equation}
with the same  $\Pi_0$(\ref{eq:origincontri}) correction.  

As the potential(\ref{linearpotapprox}) near the origin resembles the symmetric hydrogen atom, we are justified to add the same \(\Pi_{0}\)(\ref{eq:origincontri}).

Thus there are 3 nontrivial periods $\Pi_{\gamma_1}, \Pi_{\gamma_{1_-}}, \Pi_{\hat{\Gamma}}$ for the potential in Figure \ref{fig: WKBperiodsmodxEQC1}. The period structure resembles the symmetric quartic oscillator in the minimal chamber. So, the same Zinn-Justin's EQC derived in \cite{emery2021tba} will be applicable:
\begin{equation}
	\cos\left( \mathcal{B}_{med} (\Pi_{\gamma_1}) / \hbar   \right) \pm \frac{1}{\sqrt{1 + \exp[- \frac{i} { \hbar} \Pi_{\hat {\Gamma}} ]  }} = 0.
\end{equation}
Substituting for $\Pi_{\hat {\Gamma}}$(\ref{eq:closedcycle}) with $\Pi_0$(\ref{eq:origincontri}) we get: 
\begin{equation}
	\cos\left( \mathcal{B}_{med} (\Pi_{\gamma_1}) / \hbar)   \right) \pm \frac{1}{\sqrt{1 + \exp[- \frac{i} { \hbar }2 \Pi_{\hat {\gamma}} + 2\pi]  }} = 0.
 \label{eq:eqclinearqho}
\end{equation}
Now we need to fix $\pm$ sign in the EQC.

Recall for the polynomial potentials, which are regular at the origin,  $\mathcal{B}_{med} (\Pi_{\gamma_1})$ obeys QHO quantization condition 
when $\exp[- \frac{i} { \hbar} \Pi_{\hat {\Gamma}}]\rightarrow 0$. This  fixes the sign as $+$ in the EQC
(\ref{eq:eqclinearqho}) for quartic polynomial potential.

However in our case with the singularity at the origin, $\mathcal{B}_{med} (\Pi_{\gamma_1})$ obeys quantization condition(\ref{eq:eqchydro})
when $\exp[- \frac{i }{ \hbar} \Pi_{\hat {\Gamma}}]\rightarrow 0$. This requires the minus sign in (\ref{eq:eqclinearqho}). Hence the EQC to determine Voros spectrum for potential (\ref{linearpotapprox}) is
\begin{equation}
\cos\left( \mathcal{B}_{med} (\Pi_{\gamma_1}) / \hbar)   \right) - \frac{1}{\sqrt{1 + \exp[- \frac{i} { \hbar }2 \Pi_{\hat {\gamma}} + 2\pi]  }} = 0. \label{eq:eqcfinal}
\end{equation}
Solving the above EQC for $E=1,u_2=10^{-8},l=10^{-5}$  gives the Vorus spectrum. For this choice of parameters,
\(\Pi_{\hat{\gamma}}\sim \mathcal{O}(10^{(-4)}) \) (as seen in Figure
\ref{fig: epsilonmmodx}). Hence we have neglected them in determining the Voros spectrum.  Our numerical computations is tabulated in Table \ref{tbl:|x| spectrum1} \footnote{ Mathematica code of this computation can be found on the arXiv page as an ancillary file.} and they match very well with the true spectrum of $|x|$ \cite{englert2020particle,airyschwinger}. Clearly, this validation suggests that our proposed EQC is applicable for general potentials $V(r)=r+u_2/r$.
Although we focused on the limit $l \rightarrow 0$ to reproduce spectrum for the $|x|$ potential, we could determine Voros spectrum for general potentials
$V(r)=r+u_2/r+ \hbar^2 l(l+1)/r^2$ with
non-zero centrifugal for $l >0$. It appears from our modified Bethe-like ansatz(\ref{eq:bethelike}), that the correction to the \textit{proposition 3}(\ref{eq:origincontri}) for $l>0$ is
$$\Pi^{(l)}_0= 2\pi(l+1) \hbar.$$
Hence, using our numerical code, we can obtain Voros spectrum by including the above correction to $\Pi_{\hat \Gamma}$(\ref{eq:closedcycle}) in the proposed EQC(\ref{eq:eqcfinal}).
\begin{table}
	\centering    
\begin{tabular} {|c|c|c|}
	\hline 
	\(n\) & Computed \(\theta_n	\)& True \(\theta_n\)  \\ 
	\hline
	0 & 0.02852 & 0.02792 \\ 
	\hline
	1 & 1.26107 & 1.27401 \\
	\hline   
	2 & 1.76443 & 1.76715  \\
	\hline
	3 & 2.11220 & 2.11207 \\ 
	\hline
	4 & 2.35925 & 2.35919  \\ 
	\hline
	5 & 2.56402 & 2.56272\\ 
	\hline
	6 & 2.72669 & 2.72787 \\ 
	\hline
	7 & 2.87390 & 2.87165 \\ 
	\hline
	   \end{tabular}
	\caption{The numerically computed Voros spectrum for the \(|x|\) potential with \(l = 10^{-5}, u_2 = 10^{-8}\) compared to the true spectrum  }
	\label{tbl:|x| spectrum1}
\end{table}
This elaborate exercise shows that the EQC can be constructed for the general potential $V(r)$(\ref{eq:SEtypewithpole}). In fact, 
we can choose the parameters(\ref{eq:SEtypewithpole}) so that the zeros and the turning points on the positive real line (minimal chamber). Similar to what we did for $|x|+ 1/|x|$, we go back to $x \in (-\infty,\infty)$ to draw a symmetric potential with $2s+3$ cycles. For all these potentials, 
we need to modify the EQC of the smooth polynomial potential of degree $2s+4$. Near the origin, it is only the simple pole and centrifugal term which will contribute. Hence, $\Pi_{\hat {\Gamma}}$ near the origin must include $\Pi_0^{(l)}$ correction in the EQC. For highly excited states ($\theta \rightarrow \infty$), the EQC should reduce to the Bohr quantization condition(\ref{eq:eqchydro}) applicable to the singular potentials. 

Even though the methodology is straightforward, the computation of quantum periods and Voros spectrum for higher degree polynomials gets tedious. 

\section{Conclusion}
\label{sec:discussion}
In this article, first, we reviewed Bethe-like approach for quantum harmonic oscillator (QHO) and then proposed 
a modification for the hydrogen atom pseudo-momentum(\ref{eq:bethelike-withl}). This neatly reproduced the energy spectrum and the wave functions. However, for the higher degree polynomial potentials Bethe-like ansatz fails. 

We briefly presented `Thermodynamic Bethe ansatz' (TBA) method along with exact quantization conditions (EQC) leading to the spectral solutions for smooth polynomial potentials.  Even though the generalisation of TBA equations for the potentials with simple and double poles \cite{ito2020tba} is known, the EQC has not been derived. 

We showed that $|x|$ potential can be approximated to the potential with regular singularity by taking a  suitable limit of the parameters. Taking the symmetric form of the potential (\ref{eq:SEtypewithpole}) in the coordinate $x\in (-\infty, \infty)$, there will be $2s+3$ cycles. In this article, we focused on the potential for $s=0$.  

Taking hints from our proposed Bethe-like ansatz for hydrogen atom, 
we put forth {\it proposition 4} in section (\ref{sec:EQC}), stating additional correction to the quantum period $\Pi_{\hat {\Gamma}}$(\ref{eq:closedcycle}). Further, we modified the existing quartic polynomial potential EQC, imposing Bohr-Sommerfeld quantization condition applicable for singular potentials at the origin. Our proposed EQC(\ref{eq:eqcfinal}), for the potentials with single and double pole, indeed matched very well with the true spectrum for the $|x|$ potential with appropriate choices of the parameters. 
Thus we have validated our EQC proposition for the potential (\ref{eq:SEtypewithpole}) when $s=0$.

Even though we elaborated for the $s=0$ potentials, our arguments should be generalisable  for the potentials with $s>0$ as well. This requires computation of the quantum periods \cite{ito2020tba} and the modification of the smooth polynomial potential (of degree $2s+4$) EQC \cite{emery2021tba}. 
The numerical computations do get cumbersome and we will take it up in future. Such an exercise could help us to validate the $|x|^{3}$ and other odd power Voros spectrum obtained using spectral determinant approach \cite{voros1999airy}. 

\section*{Note added}
After uploading this paper on arxiv, a recent paper \cite{Gabaieqc} was kindly brought to our notice by one of the authors of that paper. Here they have derived a EQC for potential with regular singularity, although from a different Wronskian based approach.

\section*{Acknowledgements}
We would like to thank Katsushi Ito for discussions on TBA propagator. We are grateful to Marcos Mariño for sharing his notes which turned out very valuable. PR would like to acknowledge the ICTP’s Associate programme where significant progress on this work happened during her visit as senior associate. AA would like to acknowledge IIT Bombay for supporting travel to \textit{Integrability in Gauge and String Theory \textsl{2022}} conference where parts of this work were presented. PM is supported by a scholarship from the Inlaks Shivdasani Foundation.

\appendix

 \section{Regularisation of TBA for \(|x|\) }
\label{b.Regularisation of TBA}
	  In section \ref{sec:TBAmodx}, we saw the TBA equation(\ref{eq:TBAsingledoublepole}) for a linear potential with a single and double pole :
	  \begin{equation}
		  V(r) = r - E  + \frac{u_2}{r}  +  \frac{\hbar l (l +1) }{r^2} , 
	  \end{equation}
	  which takes the form: 
		  
	   \begin{align}
		  \epsilon_{1}(\theta) &= m_1 e^{\theta} - \frac{1}{2 \pi} \int_{\mathbb{R}} \frac{\log (1 + e^{-2 \hat {\epsilon }(\theta') }  - 2 \cos (2\pi l ) e^{ - \hat{\epsilon }(\theta') } ) }{ \cosh(\theta - \theta')} \nonumber\\ 
		  \hat {\epsilon}&(\theta) = \hat{m} e^{\theta} - \frac{1}{2 \pi} \int_{\mathbb{R}} \frac{\log (1 + e^{-\epsilon_1(\theta') } )    } { \cosh(\theta - \theta')} .\label{eq:TBAeqnsmodx}
	 \end{align}
	
	  In order to reproduce the Hamiltonian for the pure \(|x|\) potential, we would like to take \(u_2 = 0 , l = 0 ~\text{or}~ -1\). However, this leads to highly singular behaivour and hence we need to carefully take the limit \(u_2 \to 0 \) and \(l \to 0\) or \(l \to -1\) where the single and double pole seemingly vanish. In this limit, 
	  \begin{align*}
		  m_1 \to \frac{4}{3} E^{3/2} , ~\hat{m} \to 0.
	  \end{align*}
	   With \(E=1\), (\ref{eq:TBAeqnsmodx}) reduces to 
	  \begin{align}
		  \epsilon_{1}(\theta) &= \frac{4}{3} e^{\theta} - \frac{1}{2 \pi} \int_{\mathbb{R}} \frac{\log (1 + e^{-2 \hat {\epsilon }(\theta') }  - 2 e^{ - \hat{\epsilon }(\theta') } ) }{ \cosh(\theta - \theta')} d\theta' \nonumber\\
		  \hat {\epsilon}&(\theta) =  - \frac{1}{2 \pi} \int_{\mathbb{R}} \frac{\log (1 + e^{-\epsilon_1(\theta') } )    } { \cosh(\theta - \theta')} d\theta'.\nonumber
	  \end{align}
	  However, this TBA system is highly singular: as \(\theta \to \infty,  \hat{\epsilon} \to 0 \) and \( \log[ (1 - e^{- \hat {\epsilon }(\theta')  }  )^2] \to -\infty \). Thus additional regularisation is needed before this TBA system can be used. Following \cite{suzuki2015elementary,fendley1999airy} the regularisation of the singular term \(\epsilon_1\) must be done by subtracting a factor of \( \log (2 \pi l)\). We see this as follows : 
	  
	  \begin{align}
		  \epsilon_{1}(\theta) & \sim -\log ( 2 \pi l e^ {-A(\theta)}  + \mathcal{O}(l^2 ) ) \nonumber\\
		  \hat{\epsilon} (\theta) &\sim -\log (1 + 2\pi l B(\theta) + \mathcal{O}(l^2) ).
    \label{eq:regularisation}
	  \end{align}
	  
	  Then expanding the TBA equations in \(l\), we have 
	  \begin{align}
		  A(\theta) - \log(2\pi l) &= \frac{4}{3} e^{\theta} - \frac{1}{2 \pi} \int_{\mathbb{R}} \frac{d\theta '}{\cosh(\theta - \theta')} \log(1 + B^2(\theta')) - \log(2 \pi l)\nonumber\\ 
		  B(\theta ) &= \frac{1}{2\pi} \int _ {\mathbb{R}} \frac{d\theta'}{\cosh(\theta - \theta')} e^{-A(\theta') }.
	  \end{align}
	  Thus we see that the divergent term in \(\epsilon_1(\theta)\) ends up cancelling on both sides, leaving us with a system of equations that is no longer divergent. This regularisation, although initially appearing to be valid only for the even states with \(l = 0\) carries through exactly in the same way and gives us the same equations 
	   if instead we choose to expand around \(l = -1\). This is due to the periodicity of the only explicit \(l\) dependence in the TBA equations is given by the \(\cos(2\pi l )\) term.

	  Further, this is the same TBA system shown in \cite{fendley1999airy} (up to a overall constant shift of \(\theta \)), which was shown to be solved by the Airy functions :
	  
	  \begin{align}
		  e^{-A(\theta )} &= -2 \pi \frac{d}{dz} Ai^2(z) \\ 
		  B(\theta) = -2 &\pi \frac{d}{dz} Ai ( e^{i \pi /3} z ) Ai(e^{-i\pi/3} z),
	  \end{align}
	  with \(z = e^{\frac{2}{3}\theta} \).
	  It was argued that \(e^{-A(\theta)}\) must be the correct spectral determinant for the problem, since it vanishes at those values of \(\theta\) where \(Ai(z) = 0\) or \(Ai'(z)=0\) which correspond to the true spectrum. To get a direct derivation of this from the TBA system, let us examine what happens to the Bohr Sommerfeld quantisation under the regularisation scheme. The shift from \(\epsilon_1\)(\ref{eq:TBAeqnsmodx}) to \(\Pi_{\gamma_1}\) involves a rotation \(\theta \to \Tilde{\theta}=\theta + i\pi/2 \). This takes the form \cite{ito2020tba}
	  \begin{align*}
		  \frac{1}{\hbar}\mathcal{B}_{med} (\Pi_{\gamma_1}) &= \frac{4}{3} e^{\Tilde{\theta}} + \frac{1}{2\pi} P \int_{\mathbb{R}} \frac{d\theta'}{\sinh(\Tilde{\theta} - \theta')   }\log(1 + e^{-2\hat{\epsilon}(\theta')}- 2\cos(2\pi l)e^{-\hat{\epsilon}(\theta')    } ) \\ 
		  \frac{1}{\hbar}\mathcal{B}_{med} &(\Pi_{\gamma_1}) = 2 \pi (n + 1/2), n = 0,1,2...
	  \end{align*}
	  However under the regularisation scheme(\ref{eq:regularisation}), we have 
	  \begin{equation}
		  \mathcal{B}_{med}(\Pi_{\gamma_1}) =  \mathcal{B}_{med} (A(\theta)) + \log(2\pi l) =  2 \pi \hbar (n + 1/2)  .
	  \end{equation}
	  This implies that the points in the spectrum \( \{ \theta_i \} \) which solve the Bohr Sommerfeld condition satisfy 
	  \begin{equation}
		  e^{-\mathcal{B}_{med}(A(\theta) ) } = \mathcal{O} (2\pi l) , 
	  \end{equation}
	  which must vanish when \(l \to 0\). Hence, \(e^{-\mathcal{B}_{med}(A(\theta) ) }\) must be the spectral determinant for the problem.

\bibliographystyle{JHEP}
 \bibliography{main}

\end{document}